\documentclass[10pt]{article}
\usepackage{latexsym}
\usepackage[dvips]{graphicx}
\usepackage{overcite}
\usepackage{amssymb}
\usepackage{amsbsy}
\usepackage{amsmath}
\usepackage{mathrsfs}
\usepackage{xr}
\usepackage{booktabs} 
\usepackage{multirow} 
\usepackage{xcolor}	  
\usepackage{mathtools}
\usepackage{graphicx}
\usepackage{tabularx}
\usepackage[labelformat=simple, font=small,labelfont=bf]{caption}
\usepackage{placeins}
\usepackage{authblk}
\graphicspath{ {./plot/} }

\DeclareRobustCommand*{\Figure}[3]{
   \begin{figure}[!htb]
   \begin{center}
   \noindent
   \includegraphics[width=#2]{#1}  
   \end{center}
   \caption{#3}
   \addtocontents{lof}{\vspace{\baselineskip}}
   \label{fig:#1}
   \end{figure}
}

\newcommand{\be}{\begin{equation}}
\newcommand{\ee}{\end{equation}}

\begin{document}
\title{Effects of Dynamic Bonds on the Kinetic Pathways of Supramolecular Diblock Copolymers Disorder-Order Transition}

\author[1]{Xiangyu Zhang\footnote{E-mail: xzhan357@jh.edu}}
\author[2]{Dong Meng}
\affil[1]{Department of Chemical and Biomolecular Engineering, John Hopkins University, Baltimore, MD 21218, United States}
\affil[2]{Biomaterials Division, Department of Molecular Pathobiology, New York University, New York, NY 10010, United States}

\maketitle

\begin{abstract}
Supramolecular block copolymers (SBC) consist of covalent polymer building blocks that are connected into well-defined architectures via supramolecular bonds. Assisted by the dynamic and reversible supramolecular interactions, it is envisaged that SBC self-assemblies may exhibit more diverse morphologies, stimuli-responsivity comparing to their covalent analogues. At the fundamental level, these features are related to the free-energy landscape of self-assemblies. It is therefore of central importance to understand the impact of dynamic/reversible bonds on the free energy landscape during structure transitions. In this study, we first conduct smart Monte Carlo simulations to compare the kinetics of the disorder-order transition (DOT) of supramolecular diblock copolymers (SDBC) to that of covalent diblock copolymers (CDBC). The structural order parameter for CDBC exhibits a fast and smooth transition process across different random number seeds and initial configurations. In contrast, the SDBC system shows more diverse transition pathways, which can be classified into three types. These results suggest that reversible supramolecular interactions complicate the pathways, and bring about various intermediate structures. Next, we apply the string method to construct the minimum free energy path of the transition, from which the transition state and the free energy barrier are evaluated. It is found that the transition free energy barrier strongly correlates with the fraction of supramolecules. By decomposing the free energy into A-B interaction energy and association energy, we found that the interplay of both two effects decide the kinetic pathway and the final equilibrium structures.

\end{abstract}

\section{Introduction}
The challenges associated with processing conventional polymers (CP), such as high viscosity and long annealing times, can be mitigated by introducing reversible bonds, such as hydrogen bonding and metal coordination, leading to the formation of supramolecular polymers (SP)\cite{sijbesma1997reversible,brunsveld2001supramolecular,de2008supramolecular,lehn2002toward, feng2007supramolecular}. The reversibility of these non-covalent bonds provides SPs with several advantages over CPs, including modularity, stimuli-responsiveness, and versatility \cite{ma2014stimuli}. As a result, SPs have found applications in a wide range of fields, such as drug delivery \cite{dong2015functional, dong2011supramolecular} and catalysts \cite{gerola2017supramolecular}. \par
Two types of SP are widely studied, fully-reversible SP and partially-reversible SP. The way of connection of fully-reversible SP is only non-covalent interactions, resulting in the low stability and impairing its application \cite{yasen2020recent}. On the contrary, partially-reversible SP (PRSP) is consisted of several covalent-blocks linked by non-covalent interactions \cite{ambade2009orthogonally,yamauchi2002combinations}. PRSP has combined both advantages of CP and SP, showing a broad range of fascinating features. The covalent-blocks can ensure the stability of the polymeric materials and the reversibility of the non-covalent bond gives rises to the enriched and stimuli-responsive morphology \cite{stuparu2012phase}. But there are many questions regarding PRSP, limiting further development of functional materials. The most fundamental one of those will be the kinetic complexity resulted by the non-covalent reactions. \par
It has been proposed that the final state of SP may be dictated by kinetics rather than thermodynamics \cite{wehner2020supramolecular}. For example, the self-assembly of supramolecule polystyrene-block-poly(4-vinylpyridine) $\text{(pentadecylphenol)}_{\text{r}}$ (PS-b-P4VP $\text{(PDP)}_{\text{r}}$) shows the dependence on kinetic pathway rather than incompatibility \cite{evans2018self}. The lamellar structure formed by supramolecule is struck in unevenly distributed period size, but it cannot reach the final thermodynamic equilibrium state \cite{evans2018self}. The reason is that the struck state is preferred or chosen by the kinetics, and the free energy cost to rearrange into thermodynamics stable state is not affordable. It can be seen that the energy barrier in the free energy landscape during SP transition separates the kinetically preferred pathway and the thermodynamic stable state. Therefore, it is essential to figure out the component contribution to the free energy barrier and identify the source of the intricacy of the SP transition process. \par
The manuscript is organized as following. The disorder-lamellar transition of CP and SP is compared. First, direct dynamics simulation is applied to scan all possible transition pathways. SP transition indeed shows a more complicated pattern. The corresponding interfacial energy and association energy are analyzed. Second, intermediate struck states during SP transition are picked to characterize the most probable transition pathway to lamellar state by using string method. The free energy along the path is calculated with restraint simulation, and the free energy is decomposed into several components to find out the origin of the complexity of SP transition process. \par
\section{System Details and Methods}
\subsection{System Descriptions}
To investigate the effect of associative interactions on the kinetic transition pathway from disordered state to lamellar state, two different systems are designed. One is composed of $n_{A}=1800$ number of A homopolymer chains and $n_{B}=1800$ number of B homopolymer chains with chain length being equal, that is $N_{A}=N_{B}=10$, and both A and B homopolymer carry one associating site at the head segment, so supramolecular A$\cdot$B complex can be formed, as it is shown in figure~\ref{fig:10_figure_1.png}. The other system has A-B diblock copolymer (DBC), A homopolymer and B homopolymer, and the fraction of A-B diblock copolymer ($f_{A \text{-} B}\equiv 2 n_{A \text{-} B}/(n_{A}+n_{B})$) is equal to the equilibrium fraction of A$\cdot$B complex ($\langle f_{A \cdot B} \rangle \equiv \langle 2 n_{A \cdot B}/(n_{A}+n_{B}) \rangle$) in supramolecular system. The illustration plot can be seen in figure~\ref{fig:10_figure_1.png}. \par

The total free energy of the system can be written as the sum of bonded energy, non-bonded energy and the association energy,
\begin{equation}
\mathcal{H}=\mathcal{H}^{nb}+\mathcal{H}^b+\mathcal{H}^a
\end{equation} 
, where $\mathcal{H}^{nb}$ is the non-bonded energy, $\mathcal{H}^b$ is the bonded energy, and $\mathcal{H}^a$ is the association energy. For covalent systems, the association energy term is equal to $0$. Bonded energy is defined as, 
\begin{equation}
\mathcal{H}^{b}=\sum_{j=1}^{\Phi_{CB}}u^{b}(\left | {\bf r}_{P,j+1} - {\bf r}_{P,j} \right |)
\end{equation}
, where $\Phi_{CB}$ is the total number of covalent bonds in the system, ${\bf r}_{P,j}$ denotes the spatial position of $j$ polymer segment, and $u^{b}$ is the bonding potential, retaining connectivity. In the study, discrete Gaussian bond potential is used,
\begin{equation} \label{eq:DGB_P}
u^{b}(\left | {\bf r}_{P,j+1} - {\bf r}_{j} \right |)=\frac{3 k_{B}T}{2a^{2}}\left | {\bf r}_{P,j+1} - {\bf r}_{P,j} \right |^{2}
\end{equation}
, with $a$ being the effective bond length, $k_{B}$ being the Boltzmann constant, and $T$ being the temperature. The total non-bonded energy is given by,
\begin{equation}
\begin{split}
\mathcal{H}^{nb} & = \sum_{i=1}^{n_A\cdot N_A}\sum_{j>i}^{n_A\cdot N_A}\int d{\bf r}\int d{\bf r}' \delta({\bf r}-{\bf r}_{A,i}) u_{AA}^{nb}({\bf r},{\bf r}')\delta({\bf r}'-{\bf r}_{A,j}) \\
& + \sum_{i=1}^{n_B\cdot N_B}\sum_{j>i}^{n_B\cdot N_B}\int d{\bf r}\int d{\bf r}' \delta({\bf r}-{\bf r}_{B,i}) u_{BB}^{nb}({\bf r},{\bf r}')\delta({\bf r}'-{\bf r}_{B,j}) \\
& +\sum_{i=1}^{n_A\cdot N_A}\sum_{j=1}^{n_B\cdot N_B}\int d{\bf r}\int d{\bf r}' \delta({\bf r}-{\bf r}_{A,i}) u_{AB}^{nb}({\bf r},{\bf r}')\delta({\bf r}'-{\bf r}_{B,j})
\end{split}
\end{equation}
, where ${\bf r}_{\alpha,k}$ represents the spatial position of the $\alpha$ type $k$ segment, and
\begin{equation}
u_{\alpha \alpha'}^{nb}({\bf r},{\bf r}') \equiv \epsilon_{\alpha\alpha'} \left( 15/2\pi \right) \left(1 - r/\sigma \right)^2
\end{equation}
for $r<\sigma$, where $\sigma$ is the unit length, or $u_{\alpha \alpha'}^{nb}({\bf r},{\bf r}') = 0$ otherwise. $\epsilon_{\alpha\alpha'}$ controls the interaction strength and has the unit of $k_{B}T$, which is defined as, 
$\epsilon_{\alpha\alpha\prime}\equiv\left\{\def\arraystretch{1.2}\begin{tabular}{@{}l@{\quad}l@{}}
  $\epsilon_{\kappa}$ & if $\alpha=\alpha\prime$ \\
  $\epsilon_{\kappa}+\epsilon_{\chi_{AB}}$ & if $\alpha \neq \alpha\prime$
\end{tabular}\right.$ . $\epsilon_{\kappa}$ is the excluded volume and $\epsilon_{\chi_{AB}}$ describes the polymer incompatibility. $\epsilon_{\kappa}$ and $\epsilon_{\chi_{AB}}$ are both constants in the study. \par
The total association energy for supramolecular systems can be expressed as,
\begin{equation} \label{eq: asso_total}
\mathcal{H}^{a}=\sum_{j=1}^{N_{asso}}u^{a}(\left | {\bf r}_{A,j} - {\bf r}_{B,j} \right |)) 
\end{equation}
, where $N_{asso}$ is the total number of associated pairs in the system, and $u^{a}(\left | {\bf r}_{A,j} - {\bf r}_{B,j} \right |))$ is,
\begin{equation} \label{eq: asso_pair}
u^{a}(\left | {\bf r}_{A,j} - {\bf r}_{B,j} \right |)) = \frac{3 k_{B}T}{2\sigma^{2}}(\left | {\bf r}_{A,j} - {\bf r}_{B,j} \right |)^{2} + h_{A} 
\end{equation}
, where $h_{A}$ is the association constant to control the equilibrium A$\cdot$B complex fraction. The detail of the implementation of association movements in Monte Carlo simulations can be found in the cited paper \cite{zhang2025investigation}. \par
 $\epsilon_{\kappa}$ (excluded volume effect) is set to be equal to $0.12$, and the incompatibility between A-type chain and B-type chain is $0.06$ ($\epsilon_{\chi_{AB}}$) at the starting point, which is the disordered state. Then, $\epsilon_{\chi_{AB}}$ is increased to $0.2$ to drive the transition to lamellar state, corresponding to the quenching process. The association energy barrier ($h_{A}$) is set as $-3.9$ for supramolecular system, and the equilibrium fraction of A$\cdot$B complex ($\langle f_{A \cdot B} \rangle$) at lamellar morphology is $0.755$. Accordingly, the fraction of A-B DBC ($f_{A \text{-} B}$) in covalent system is also $0.755$, meaning the number of A-B DBC is $n_{A \text{-} B}=1359$, and the number of both A and B homopolymer chain is $442$, corresponding to the same composition as the supramolecular system. The box size is chosen to be $16\times 16 \times 16 \ \sigma ^{3}$. To induce the lamellar direction, two hard walls with the repulsion to A-type chain are planted along y-direction. Figure~\ref{fig:10_figure_2.png} shows the example of the system transition from the disordered state to the lamellar state. Smart Monte Carlo is applied first to scan all possible pathways by changing the random number seed and the initial configuration. After knowing possible pathways, swarms of trajectories string method is used to characterize the minimum free energy pathway (MFEP) during the transition process, and the free energy difference along the path is calculated. \par

\Figure{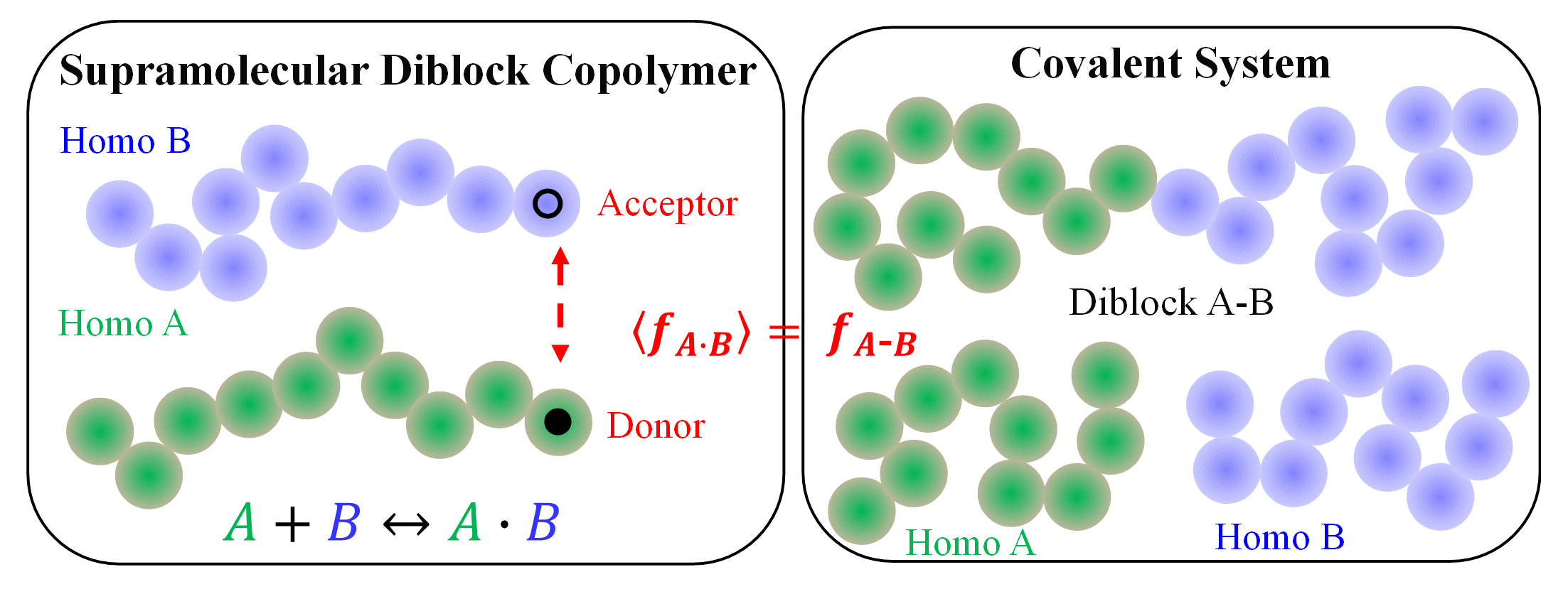}{0.98\linewidth}{Illustration plot of covalent bonded A/B/A-B system and supramolecular complex A/B/$\text{A}\cdot\text{B}$ system.}
\Figure{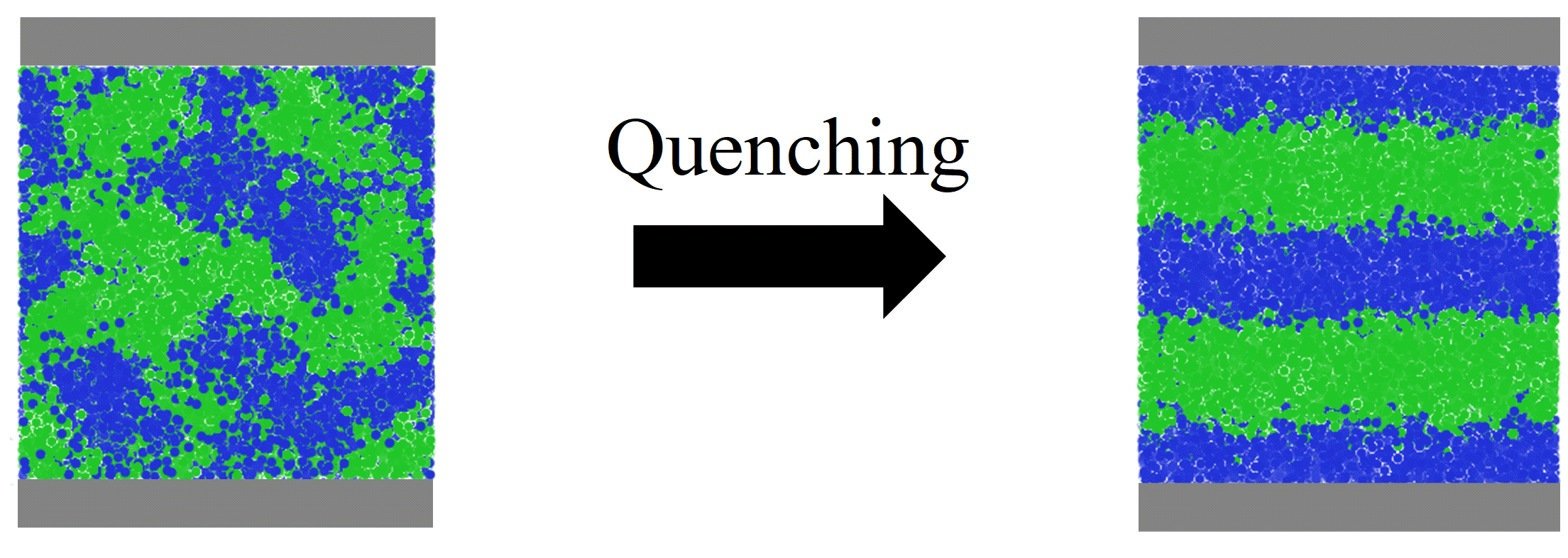}{0.9\linewidth}{The example plot of the transition of the A/B blending system from the disordered state to the lamellar state.}

\subsection{Smart Monte Carlo Simulation}
Smart Monte Carlo includes the stochastic and systematic effects for the hopping trial \cite{allen2017computer}. And it has been examined to show the ability to capture the dynamics property for polymer melt system \cite{muller2008single}. Therefore, smart Monte Carlo trial moves are implemented to search for the possible dynamics path. The trial displacement from "o" state to "n" state of the segment "i" is written as,
\begin{equation} \label{10result_SMCT}
\delta {\bf r}= {\bf F}_{i,o}({\bf r}) \Delta A + \Delta {\bf r}^{G}
\end{equation}
where ${\bf F}_{i,o}({\bf r})$ denotes the covalent bonded-force in the old position, $\Delta A$ is an input parameter, and $\Delta {\bf r}^{G}$ is random number which obeys the Gaussian distribution, with the mean being zero and the variance being $2A$ in each direction \cite{rossky1978brownian}. And box-muller transformation is used to convert uniform distributed random number to normal distributed random number \cite{box1958note}. Thus, the Gaussian random variable with mean being zero and the variance being $2A$ can be obtained by multiplying normal distributed random number with $\sqrt{2A}$. To fulfill the detailed balance, the acceptance criterion can be derived as,
\begin{equation}
acc(old \to new)=\min{(1, \exp{[-\beta(\mathcal{H}_{n}-\mathcal{H}_{o}+\frac{{\bf F}_{o}({\bf r})+{\bf F}_{n}({\bf r}')}{2} \delta {\bf r} +\frac{\Delta A}{4}( {\bf F}_{n}({\bf r}') - {\bf F}_{o}({\bf r}) ) )]}	)}
\end{equation}
where $\mathcal{H}_{n}$ and $\mathcal{H}_{o}$ are the energy in the new and old state, ${\bf r}'$ is the new position after the displacement, ${\bf F}_{n}({\bf r}')$ denotes the covalent bonded-force in the new state \cite{muller2008single}. \par

\subsection{Swarms of Trajectories String Method}\label{sec:10_result_STSM}
The Cartesian coordinates sets of the system is ${\bf x}\in \mathbb{R}^{3N}$. Correspondingly, the set of collective variables (CV) to map system's configuration to the reaction coordinates for image ID "$s$" can be expressed as, $\tilde{{\bf z}}({\bf x})=\tilde{z}_{1}({\bf x}), \tilde{z}_{2}({\bf x}),..., \tilde{z}_{N}({\bf x})$. The reaction coordinates chosen for the calculation is the particle density field ($\phi({\bf c})$) in this study. The simulation box is divided into $16\times16\times16$ cubic cells with each one being $1 \sigma^{3}$. To convert the particle coordinates to the field distribution, clouds in cells method is used, in which each particle is treated as one cube with the length of each side being $1 \sigma$ and the volume being $1 \sigma^{3}$. \par
The method implementation basically follows the procedure in the cited paper \cite{pan2008finding}, and the smart Monte Carlo hopping movement and the association trials are implemented. The potential of the mean force of the collective variables can be written as,
\begin{equation}
e^{-\beta W({\bf z})} = \frac{\int d{\bf x} \delta(\tilde{{\bf z}}({\bf x}) - {\bf z})e^{-\beta \mathcal{H}({\bf x})}}{\int d{\bf x} e^{-\beta \mathcal{H}({\bf x})}} . 
\end{equation}
The evolution of the CV over time step $\delta \tau$ according to non-inertial Brownian dynamics is,  
\begin{equation} \label{eq:10result_BD}
\tilde{z}_{i}(\delta \tau) = \tilde{z}_{i}(0) + \sum_{j}(\beta D_{ij}[\tilde{\bf z}(0)]F_{j}[\tilde{\bf z}(0)]+\frac{\partial D_{ij}[\tilde{\bf z}(0)]}{\partial {\bf z}_{j}}) \delta \tau + R_{i}(0)
\end{equation}
where $D_{ij}$ is the diffusion tensor, $F=-\triangledown W({\bf z})$ is the mean force, and $R_{i}(0)$ is the Gaussian random variable, with the mean value being zero and the variance being $2D_{ij}\delta \tau$ \cite{ermak1978brownian}. If we want to find the MFEP for two stable basins, the last term of eq. \ref{eq:10result_BD} ($R_{i}(0)$) has to be zero, because the mean of the Gaussian variable is zero, which is the most probable value \cite{pan2008finding}. The eq. \ref{eq:10result_BD} becomes, 
\begin{equation} \label{eq:10result_BDMFEP}
\tilde{z}_{i}(\delta \tau) = \tilde{z}_{i}(0) + \sum_{j}(\beta D_{ij}[\tilde{\bf z}(0)]F_{j}[\tilde{\bf z}(0)]+\frac{\partial D_{ij}[\tilde{\bf z}(0)]}{\partial {\bf z}_{j}}) \delta \tau .
\end{equation}
And it also has to satisfy the condition, $({\bf D}{\bf F}+\frac{\partial D_{ij}[\tilde{\bf z}(0)]}{\partial {\bf z}_{j}})^{\perp}=0$, because tension tangent to the MFEP should be zero. Thus, it can be found that $(\langle\Delta \tilde{{\bf z}}\rangle)^{\perp}=0$ for a converged curve, and $ \Delta \tilde{{\bf z}} \equiv \tilde{\bf z}(\delta \tau) - \tilde{\bf z}(0)$. Finally, the update of CV in each image for one iteration step can be written as, 
\begin{equation}\label{eq:10result_CVE}
\tilde{\bf z}(\delta \tau) = \tilde{\bf z}(0) + \frac{1}{S}\sum_{k=1}^{S}( \tilde{\bf z}( {\bf x}^{k}(\delta \tau)) - \tilde{\bf z}(0))
\end{equation}
where $S$ is the number of trajectories used for each step \cite{maragliano2014comparison}. After updating CV, it is needed to redistribute each image to ensure the equal-distance distribution for all of image points along the path. Because CV will finally reach the stable basin without any external restrictions due to the free energy gradient. So, reparametrization step is implemented, which redistributes each point with equal Euclidean distance based on the piece linear interpolation method \cite{maragliano2006string}. In general, the procedure to perform swarms of trajectories string simulation can be organized into following steps: (1) Build up the initial path (in our algorithm, linear initialization is used); (2) Create the particle configuration compatible with the target density field by using the restraint simulation; (3) Run short swarms for all trajectories; (4) Evolve CV according to the eq. \ref{eq:10result_CVE}; (5) Redistribute all image points to guarantee the equal distribution, that is reparametrization.  \par

\subsection{Free Energy Calculation along the Path}\label{sec:10result_FEC}
The minimum free energy pathway is obtained by using the method discussed in section~\ref{sec:10_result_STSM}. Next, restraint simulation is running to characterize the free energy along the path, the procedure follows the method described in the cited paper \cite{muller2012transition}. The umbrella potential is used to restrain the fluctuation of the CV ($\tilde{\bf z}(\bf x)$) from its converged value, which is also the restraint value (${\bf z}_{c}$),
\begin{equation}
\beta\mathcal{H}_{c}({\bf x}\vert {\bf z}_{c})=\beta \mathcal{H}({\bf x})+\int d{\bf x} \frac{\lambda}{2}[{\bf z}_{c} - \tilde{\bf z}(\bf x)]^{2}  . 
\end{equation}
If $\lambda \to \infty$, the free energy of the restraint system will recover the free energy of the free system with CV being ${\bf z}_{c}$, that is $\beta F_{c}[{\bf z}_{c}]\equiv - \ln{\int d {\bf x}\exp{(-\beta\mathcal{H}_{c}({\bf x}\vert {\bf z}_{c}))}}$. The chemical potential can be obtained by calculating the fluctuation of the restraint system,
\begin{equation}\label{eq:10_result_mu}
\frac{\delta F_{c}[{\bf z}_{c}]}{\delta {\bf z}_{c}}=\lambda k_{B}T[{\bf z}_{c} - \langle \tilde{\bf z}({\bf x}) \rangle_{c}] \xrightarrow{\lambda \to \infty} \mu({\bf x} \vert {\bf z}_{c})  . 
\end{equation} 
The free energy along the path can be computed by substituting the chemical potential calculated from eq. \ref{eq:10_result_mu} into the following equation, 
\begin{equation}
\frac{d \mathcal{F}[{\bf z}_{c,s}]}{ds}=\int d{\bf x}	 \frac{\partial {\bf z}_{c,s}({\bf x})}{\partial s} \mu({\bf x} \vert {\bf z}_{c,s})
\end{equation} 
where ${\bf z}_{c,s}$ denotes the converged CV of image "$s$".  \par
\section{Results and Discussions}
\subsection{Direct Dynamics Simulation}
Smart Monte Carlo simulation is implemented first. Several different random number seeds and initial configurations at disordered state are used to ensure that all possible transition pathway will be scanned. Order parameter ($S(q)$) is used to indicate the system structure. The way to calculate $S(q)$ is defined as following,
\begin{equation}
\begin{split}
& S(q)  \equiv \\ 
& [(\sum_{j=1}^{n_{A}N_{A}} \cos{({\bf q} \cdot {\bf r}_{A,j})} - \sum_{j=1}^{n_{B}N_{B}} \cos{({\bf q} \cdot {\bf r}_{B,j})})^{2} + (\sum_{j=1}^{n_{A}N_{A}} \sin{({\bf q} \cdot {\bf r}_{A,j})} - \sum_{j=1}^{n_{B}N_{B}} \sin{({\bf q} \cdot {\bf r}_{B,j})})^{2}]/N_{T}
\end{split}
\end{equation} 
where $N_{T}$ is the total number of segments in the system. The wave vector length ($\left\| q^{*} \right\|$) used in the calculation is chosen to be $2\pi/L_{0}$, in which $L_{0}$ is the lamellar period. The result of order parameter as a function of MC steps is shown in figure~\ref{fig:10_figure_3.png}. The configuration at $S(q^{*}) \to 0$ is identified to be disordered state, and the characteristic value for lamellar state is found to be $S(q^{*})\approx 14000$. In covalent system, that is figure~\ref{fig:10_figure_3.png} (a), the order parameter behaviors of all of systems with different initial configurations and random number seeds show the similar pattern. The transition process is rapid and smooth. The only slight difference among all covalent systems is the ascending slopes, suggesting the less complexity in the kinetic transition pathway of covalent system. But the behavior in supramolecular system is quite special. First, let us examine these red curves in figure~\ref{fig:10_figure_3.png} (b), named path type-I for the convenience of discussion. They show covalent-like transition pathway, which is fast and smooth. The blue curves exhibit the transient plateau before eventually reaching lamellar state, named path type-II. The intermediate structure is shown in the snapshot below the figure. At $S(q^{*})$ equal to around $8000$, layers are connected by several tunnels, transporting the particles to reach the mass equilibrium. Similarly, the tunnel structure can also be observed at $S(q^{*})=10000$ plateau. The green curves show the persistent plateau at order parameter being around $4000$ and it will not transit to lamellar structure, named path type-III. The snapshot below suggests that it is still in disordered state, but it seems to have larger degree of segregation. The appearance of three different types of transition paths signal the kinetic pathway complexity of the supramolecular system in comparison to the covalent system. The understanding of this complexity can be provided by analyzing the AB interaction energy and fraction of $\text{A}\cdot\text{B}$ complex. \par

\Figure{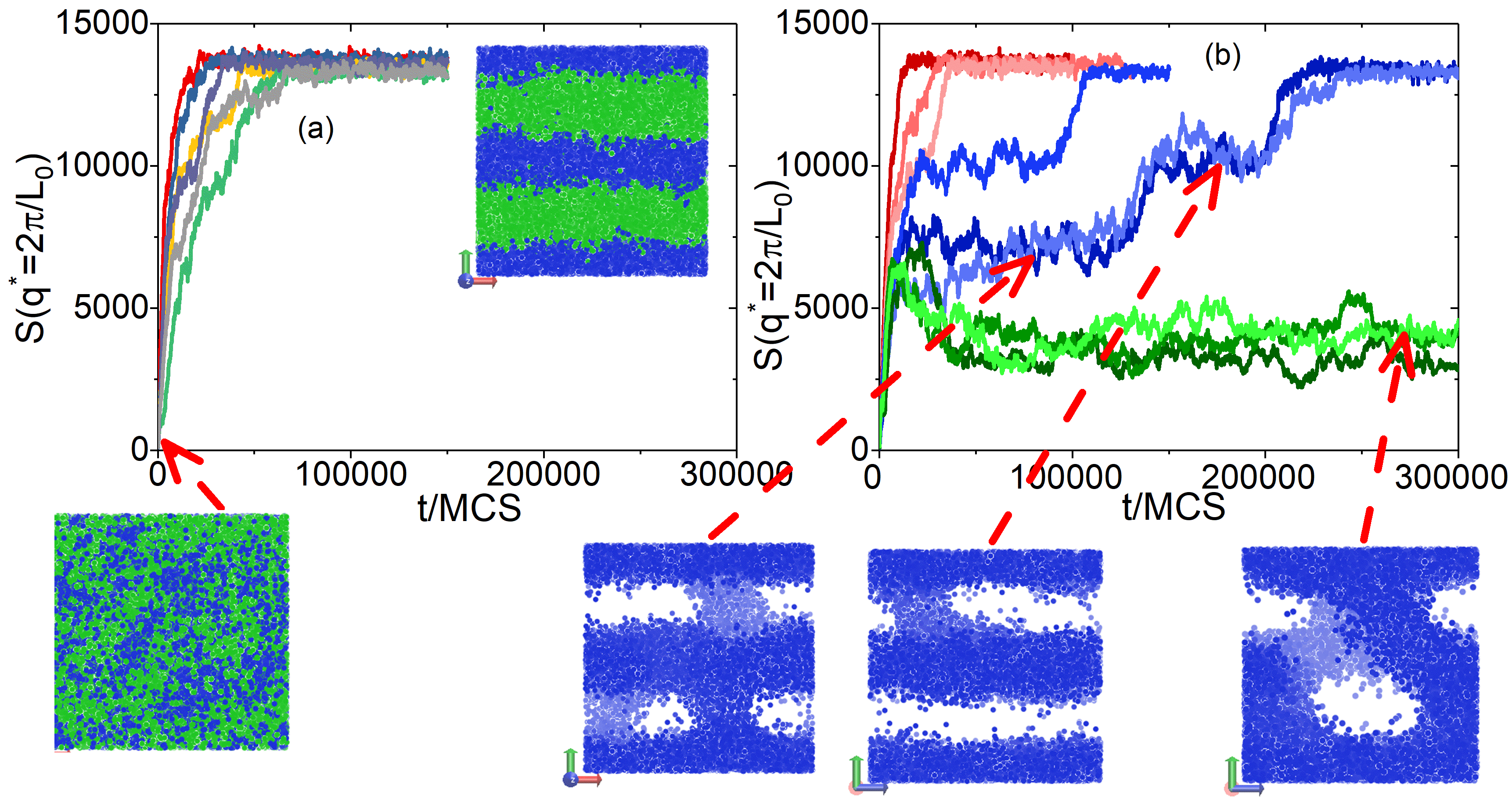}{0.98\linewidth}{Order parameter with wave vector length being $2\pi/L_{0}$ plotted against MC steps for (a) covalent system and (b) supramolecular system, and the corresponding snapshot at a certain $S(q^{*})$ value is also shown in the inset.}

We start by examining the change of AB interaction energy ($E_{AB}$) along three types of transition paths in figure~\ref{fig:10_figure_4.png} (a). Along path type-I, $E_{AB}$ quickly reaches the equilibrium value, which is indicated by the black dashed line in the plot. Along the type-II path, before reaching the lamellar phase, $E_{AB}$ dips slightly below the equilibrium value. The difference is small but appreciable, suggesting that the intermediate structures observed along path II have the local structure similar to that of the lamellar phase, with well defined A and B domains and an AB interfacial area per volume very similar to that of the lamellar phase. The more pronounced changes in $S(q^{*})$ plots as system transition to lamellar phase is a result of improved structural ordering at a longer length scale (at the length scale of lamellar period). This can also be signified by the snapshot shown in figure~\ref{fig:10_figure_3.png}. Along the type-III path, $E_{AB}$ is found to be significantly lower than the equilibrium value, indicating a long-lived morphology with a higher degree of A-B segregation than the lamellar phase. To further understand the origin of these differences, we plot the fraction of $\text{A}\cdot\text{B}$ diblock complex in the supramolecular system as the function of Monte Carlo steps in figure~\ref{fig:10_figure_4.png} (b). Along path type-I, the fraction of $\text{A}\cdot\text{B}$ diblock complex quickly reaches the equilibrium value, (indicated by the black dashed line). Along Path type-II, before reaching lamellar phase, the fraction stays appreciably smaller than the equilibrium value. A significantly lower fraction is observed along path type-III that stays as such throughout the simulation. This analysis suggests a correlation between the prolonged plateaus seen in $S(q^{*})$ plots with the lower AB interaction energy as well as lower fraction of $\text{A}\cdot\text{B}$ complex in the supramolecular system during the phase transition. The above results suggest that both degree of AB segregation and the system composition fluctuation, that is the fraction of $\text{A}\cdot\text{B}$ complex, play the important role in deciding transition pathway. On the other hand, the covalent system only involves the extent of AB contact area, simplifying the transition process. Therefore, we need to characterize the transition pathway on the ground of thermodynamics by using string method. \par

\Figure{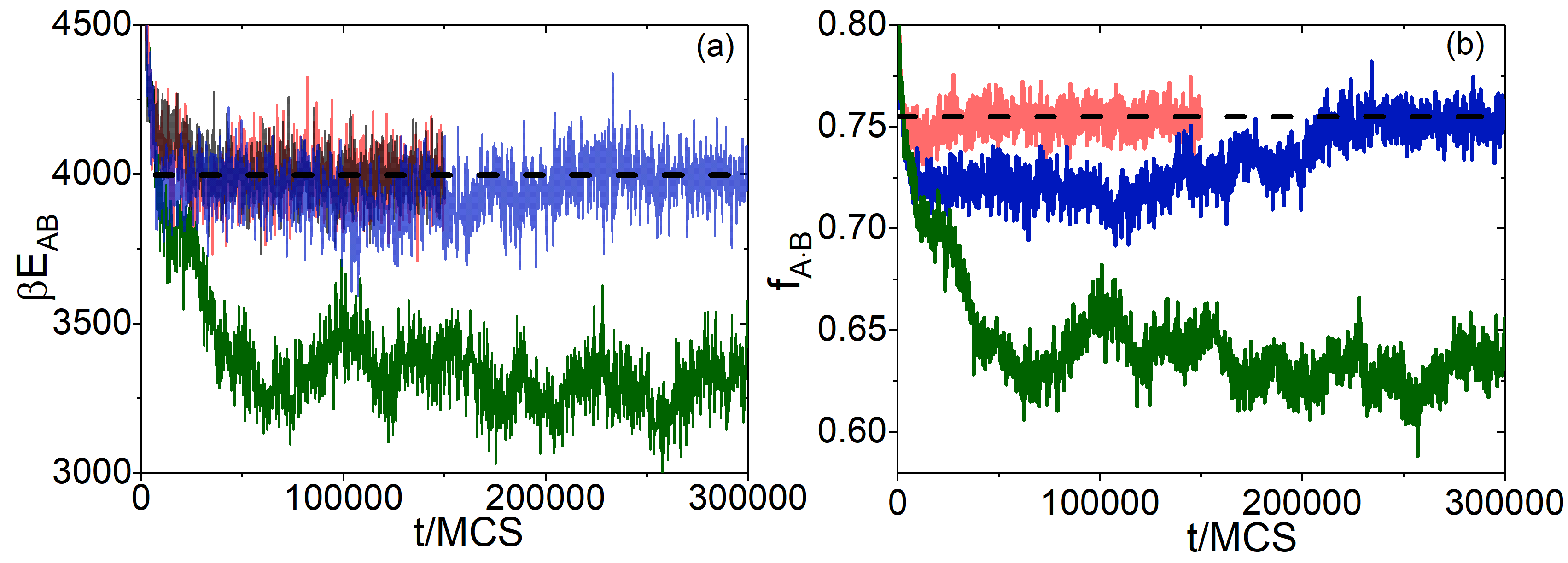}{0.98\linewidth}{(a) A-B interfacial energy and (b) fraction of A-B complex plotted against the MC steps in supramolecular system. The equilibrium value at lamellar state in both supramolecular and covalent system is indicated by black dashed line in the plot.}

\subsection{Most Probable Path Characterized by String Method}
Two most interesting path behaviors will be the transient plateau observed in path type-II and persistent plateau observed in path type-III. So, the corresponding configuration is picked from direct dynamics simulation and chosen to be the terminal point of the path. Accordingly, the starting point of the path is the equilibrium lamellar structure. \par
The free energy along the most probable transition pathway connecting type-III persistent plateau to the lamellar structure is plotted in figure~\ref{fig:10_figure_5.png}. $d \mathcal{F}/ds$ calculation follows the method described in the section~\ref{sec:10result_FEC}. Accordingly, the free energy can be obtained by doing the integral. The end point at the left side suggests the lamellar state, and the right end point corresponds to the stuck plateau structure. The whole free energy curve is shifted to make the free energy at stable lamellar state be $0$. The reason why two ends points are not at the stable basin is that those two points are taken directly from smart MC simulation, and those configurations may not fall at the minimum point in the basin due to fluctuations in the simulation. The struck structure indeed has the lower free energy than the lamellar state, indicating the more stable configuration. That is why we can observe persistent plateau in figure~\ref{fig:10_figure_3.png} (b). So, several direct dynamics simulations starting from lamellar state are running to see whether the transition to struck plateau state can be observed. It is found that the system morphology does not do any changes even with millions of MC steps, still staying at lamellar state. The reason is the high energy barrier during the transition. The system needs to break the lamellar layer and form the tunnels connecting layers, like showing in the snapshot, to transit to the struck state, but the energy penalty is too high. Therefore, the system cannot cross that barrier, and can only stay at the lamellar structure. To obtain more understanding about why the struck state is more stable, free energy is decomposed into several components. \par

\Figure{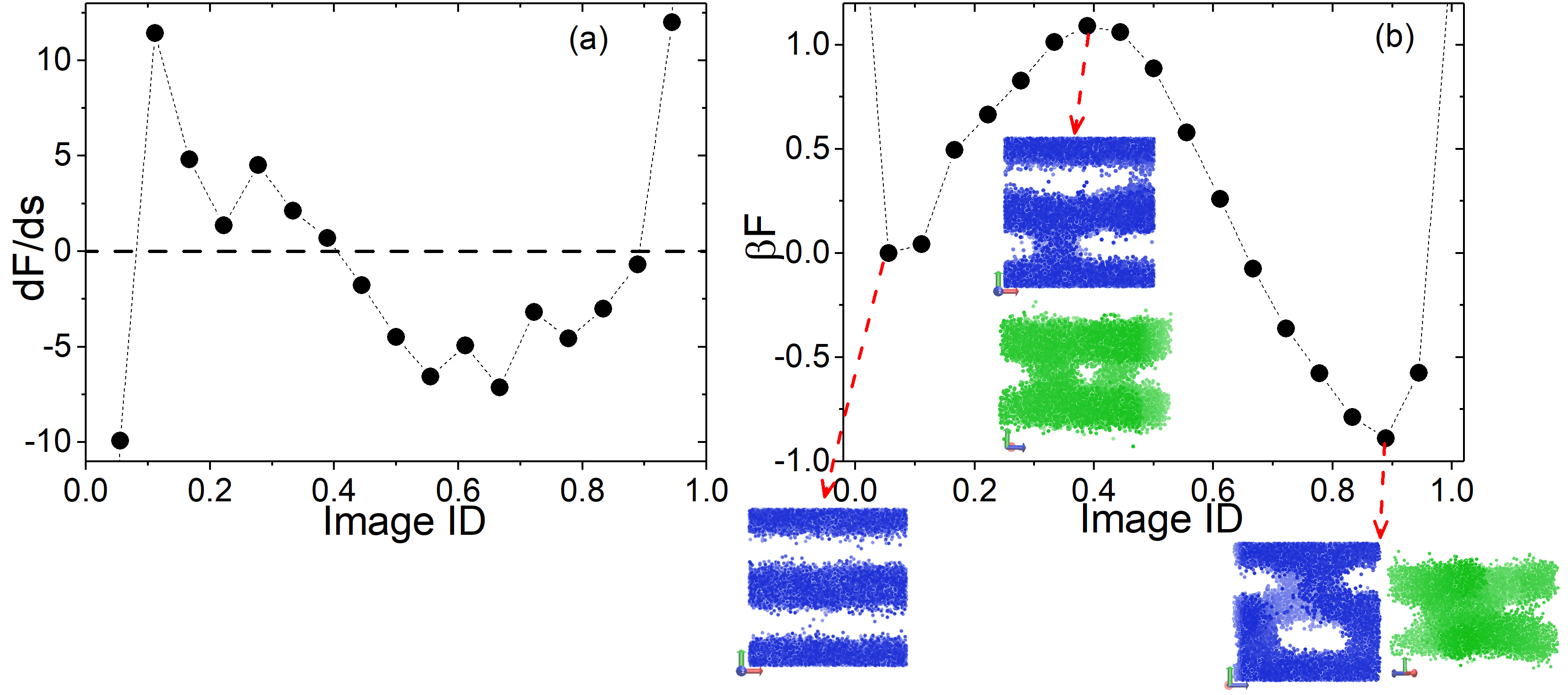}{0.98\linewidth}{The (a) free energy derivative and (b) free energy along the most probable transition pathway connecting persistent plateau in path type-III and the lamellar structure. The snapshots from simulation are shown in the inset.}
The AB interaction energy and the association energy are plotted in figure~\ref{fig:10_figure_6.png}. The AB interaction energy and association energy show the opposite trend. The decrease of AB interaction suggests the less contact area between A and B polymers, and indicates the larger degree of segregation. On the contrary, the increase of the association energy signifies the lower degree of $\text{A}\cdot\text{B}$ complexation. The magnitude of the AB interaction energy decrease is around $0.2$, but the magnitude of association energy increase is close to $0.15$. Hence, the decrease of AB interaction energy can compensate for the association energy gain, stabilizing the struck plateau structure. The similar behaviors are also observed in the brutal force simulation results in figure~\ref{fig:10_figure_4.png}. The above results indicate the competition between degree of "complexation", implied by association energy, and extent of "segregation", suggested by AB interaction energy, in supramolecular system. The interplay of both effects together decides the kinetics pathway and the final reached structure. Conversely, the composition is fixed in the covalent system, so, only one type of transition pattern can be observed. \par

\Figure{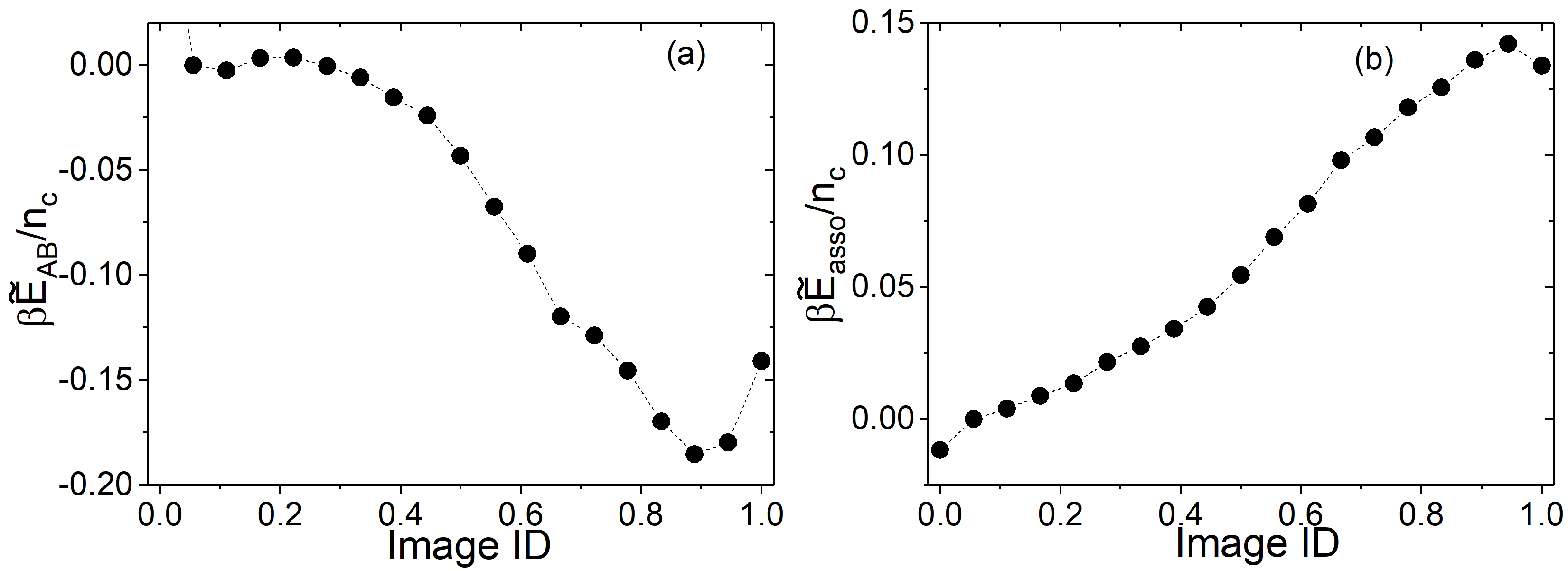}{0.98\linewidth}{The (a) AB interaction energy and (b) association energy both normalized by number of chains in the system along the most probable transition pathway connecting persistent plateau in path type-III and the lamellar structure.}
Next, a configuration at transient plateau ($S(q^{*})\approx8000$) from type-II path is picked and set as the terminal point of the path calculation. The free energy result along the transition pathway is shown in figure~\ref{fig:10_figure_7.png}. It can be seen that there is indeed a meta-stable basin for transition plateau structure. Correspondingly, in free energy derivative plot, three points at around ID image being $0.8$ are below $0$. It tells that the transient plateau observed in brutal force simulation in figure~\ref{fig:10_figure_3.png} (b) comes from this local minimum point. The system would rather stay here for a while due to the energy barrier. The degree of energy barrier is relatively low in this path, so, it will eventually reach the lamellar state. And that is also what we observed in direct dynamics simulation results. The free energy is also decomposed into AB interaction energy and association energy in figure~\ref{fig:10_figure_8.png}. The value of AB interaction energy decrease at transient plateau structure (image ID being $0.8$) is around $-0.01$. But the association energy gain here is more than $0.03$. Therefore, the increase of association energy gain is higher than the loss of AB interaction energy at transient plateau structure. That is why this point is a local minimum or meta-stable point. Additionally, some points in AB interaction energy plot at image ID between $0.4$ to $0.6$ are not so smooth. The reason is that if we look at figure~\ref{fig:10_figure_3.png} (b), there is another plateau at around $S(q^{*})\approx10000$, suggesting another meta-stable point. Also, one point in free energy derivative plot in figure~\ref{fig:10_figure_7.png} touches $0$ with image ID being $0.4$, indicating the existence of local basin. But the free energy does not capture that point due to limited resolutions. Overall, the qualitative feature is very clear. The occurrence of the plateau observed in direct dynamics simulations indeed suggests the existence of the meta-stable point. And the plateau length scale can be perfectly explained by the free energy path calculation. In further, the role of each free energy component playing in the transition path is figured out. \par

\Figure{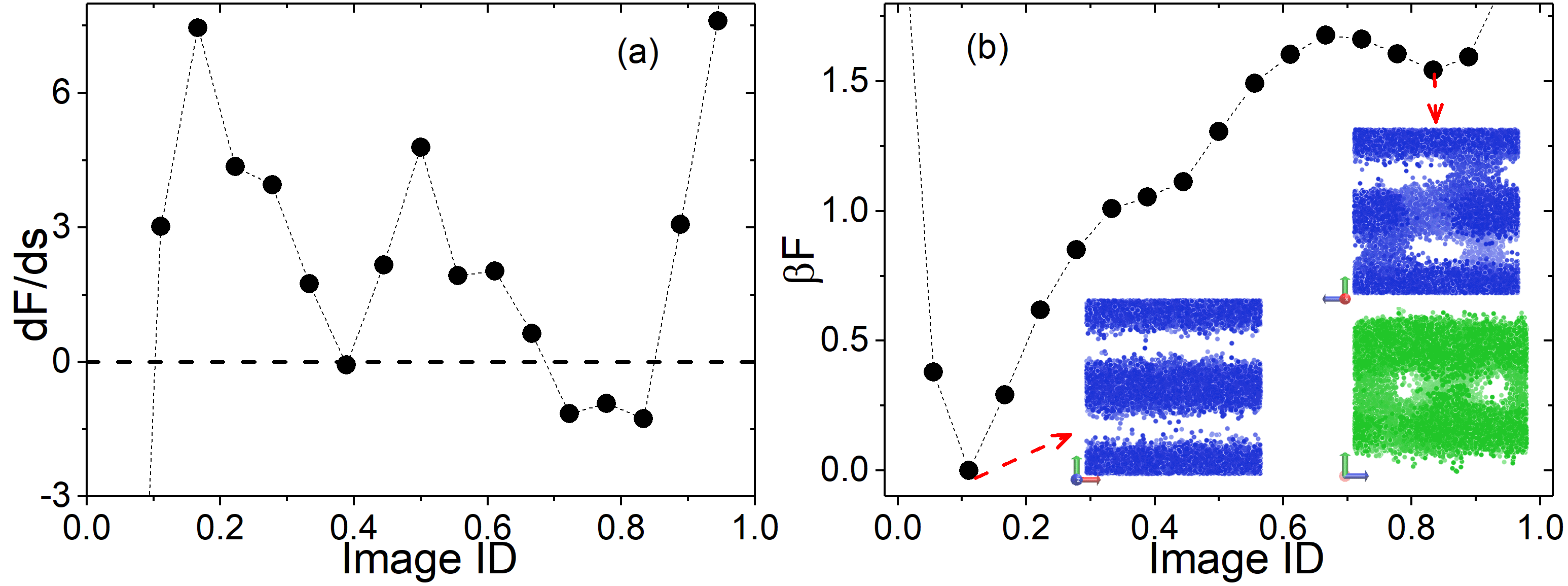}{0.98\linewidth}{The (a) free energy derivative and (b) free energy along the most probable transition pathway connecting transient plateau ($S(q^{*})\approx8000$) in path type-II and the lamellar structure. The snapshots from simulation are shown in the inset.}

\Figure{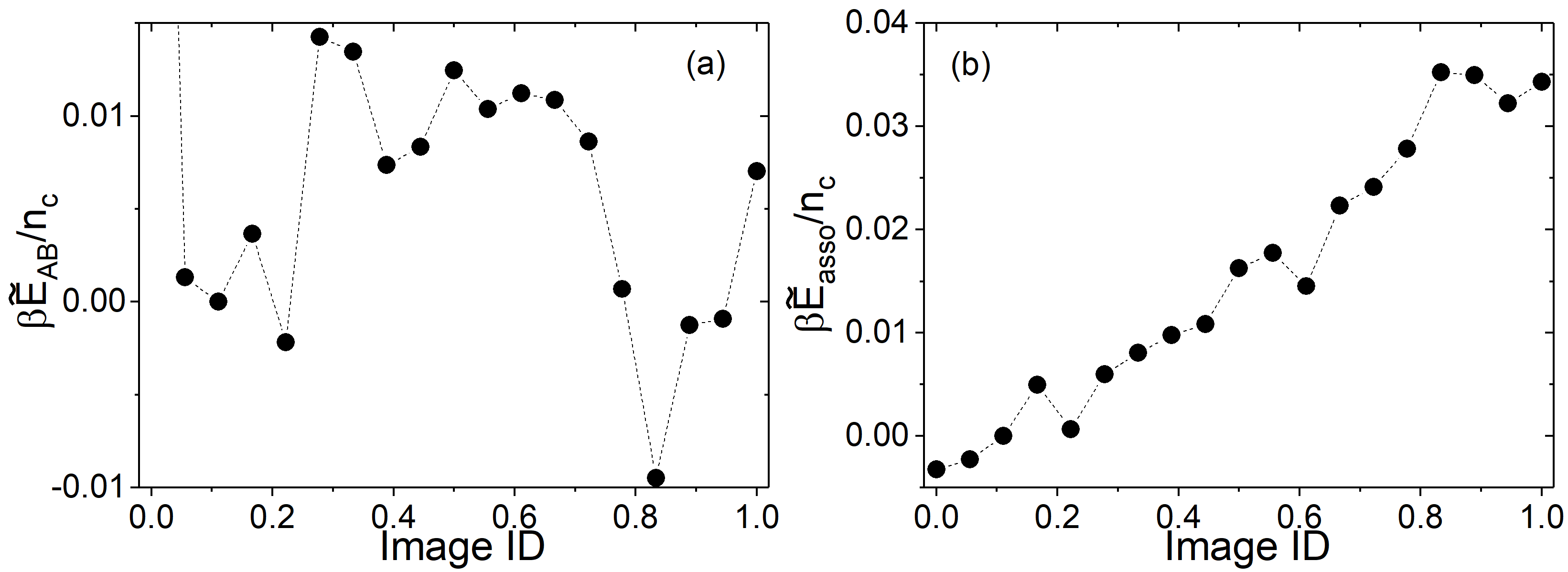}{0.98\linewidth}{The (a) AB interaction energy and (b) association energy both normalized by number of chains in the system along the most probable transition pathway connecting transient plateau in path type-II and the lamellar structure.}

\section{Conclusion}
Direct dynamics simulations (Smart MC) are first performed to explore all possible transition pathways. To understand the origin of different pathway types, the string method is implemented to characterize the thermodynamic properties. The results reveal that the supramolecular system exhibits increased kinetic pathway complexity during the disorder-to-lamellar transition. By decomposing the free energy into supramolecular association energy and AB interaction energy, the source of this complexity is identified, confirming the crucial role of their interplay in determining the kinetic pathways. In contrast, the covalent system follows a single pathway pattern due to the absence of compositional fluctuations, as the fraction of A-B DBC remains fixed. The observations from direct dynamics simulations are well-explained by the free energy path calculations, which further show that the plateau length scale correlates with the magnitude of the energy barrier along the pathway. This study highlights the critical role of reversible supramolecular interactions in determining transition pathways and provides fundamental insights into the self-assembly process of supramolecular systems. \par

\newpage
\bibliographystyle{unsrt}
\bibliography{citation}

\end{document}